\documentclass[journal=nanoletters,manuscript=letter]{achemso}
\usepackage[latin9,utf8]{inputenc}
\usepackage{float}
\usepackage{textcomp}
\usepackage{amsmath,amssymb,amscd,hyperref}
\usepackage{graphicx}
\makeatletter
\usepackage{url}
\usepackage{grffile}
\usepackage{bbold}
\usepackage{comment}
\usepackage{dirtytalk}
\usepackage{chngcntr}
\usepackage{tabularx}
\usepackage{multirow}
\usepackage{soul}
\usepackage{cleveref}
\usepackage{braket}
\DeclareUnicodeCharacter{2212}{-}
\DeclareUnicodeCharacter{0301}{\hspace{-1ex}\'{ }}

\newcommand{\keyword}[1]{\noindent\textbf{Keywords:} #1}

\usepackage{dcolumn}
\usepackage{color}
\DeclareGraphicsExtensions{.png .jpg .pdf}
\hypersetup{
     colorlinks = true,
     linkcolor = blue,
     anchorcolor = blue,
     citecolor = blue,
     filecolor = blue,
     urlcolor = blue
     }
\usepackage{braket}
\usepackage{physics}
\usepackage{array}
\usepackage{booktabs}
\usepackage{soul}

\makeatother

\author{X. D. Wang}
\altaffiliation{These authors contributed equally to this work.}
\affiliation{School of Materials and Physics, China University of Mining and Technology, Xuzhou 221116, P. R. China}

\author{J. F. Oliveira da Silva}
\altaffiliation{These authors contributed equally to this work.}
\affiliation{COMMIT, Department of Physics, University of Antwerp, Groenenborgerlaan 171, 2020 Antwerp, Belgium}

\author{Z. H. Tao}
\affiliation{COMMIT, Department of Physics, University of Antwerp, Groenenborgerlaan 171, 2020 Antwerp, Belgium}
\altaffiliation{These authors contributed equally to this work.}

\author{H. M. Dong}
\email{hmdong@cumt.edu.cn}
\affiliation{School of Materials and Physics, China University of Mining and Technology, Xuzhou 221116, P. R. China}

\author{K. Chang}
\email{kchang@zju.edu.cn}
\affiliation{Center for Quantum Matter, School of Physics, Zhejiang University, Hangzhou 310027, P. R. China}

\author{M. V. Milo\v{s}evi\'{c}}
\email{milorad.milosevic@uantwerpen.be}
\affiliation{COMMIT, Department of Physics, University of Antwerp, Groenenborgerlaan 171, 2020 Antwerp, Belgium}

\title{Symmetry-Tunable Skyrmions and Merons in Magnetic Nanodisks via Spatially Engineered Anisotropy}

\date{\today}
\begin{document}

\date{\today}

\begin{abstract}
We demonstrate that spatially engineered magnetic anisotropy can stabilize skyrmion and meron spin textures in magnetic nanodisks even in the absence of Dzyaloshinskii-Moriya interaction (DMI). Using a constrained analytical model and unconstrained micromagnetic simulations, we show that competing perpendicular and in-plane anisotropies can generate non-collinear topological textures in non-chiral magnetic systems. We further show that DMI and dipolar interactions lift the helicity degeneracy and select preferred chiral configurations; micromagnetic simulations were used to identify physically stable states. These results establish anisotropy-patterned nanodisks as a platform for studying DMI-free topological spin textures and their controllable magnetic response. We also show that arrays of anisotropy-engineered skyrmions can control spin-wave transmission by manipulating their vorticity arrangement, pointing to reconfigurable magnonic elements based on non-chiral topological textures.
\end{abstract}
\keyword{Skymions, Magnetic Nanodisks, Spin textures, Symmetry-tunable}
\maketitle

%\section{Introduction}
Topological spin textures, such as magnetic skyrmions and merons, have garnered intense interest over the past decade due to their particle-like stability, nanoscale dimensions, and potential as information carriers in next-generation spintronic devices \cite{nagaosa_2013, fert2017a}. Their inherent topological protection offers resilience against local perturbations, making them attractive for low-power memory, logic, and neuromorphic computing applications \cite{DaC2025}. Despite these promising attributes, achieving tunable skyrmions and their lattice structures in non-chiral conventional magnetic materials remains challenging. Conventional skyrmions stabilized by chiral DMI have chirality determined by the DMI sign and symmetry, thereby limiting their internal degrees of freedom compared with systems featuring designed anisotropy landscapes \cite{yu_voltage2024}. Furthermore, the reliance on chiral exchange interactions, absent in many technologically relevant centrosymmetric magnets, imposes severe material constraints. Consequently, it is important to determine whether skyrmionic textures can be created and modified in non-chiral systems, and whether their symmetry and functional properties can be engineered.

Recent efforts have explored geometric confinement and dipolar frustration as alternative mechanisms for stabilization \cite{Gilbert2015, yoshimochi2024a}. However, the control over axial symmetry and topological charge without DMI, key degrees of freedom that distinguish distinct spin states, has remained elusive. Prior demonstrations often yield fixed, unidirectional textures or require complex field protocols, precluding deterministic device integration \cite{Li2014, Zhao2022nl}. Proposals for skyrmion stabilization without DMI have been explored through frustrated exchange interactions and dipolar fields \cite{hou2017}. Here, we propose a different approach based on spatially patterned competing anisotropies, specifically an in-plane uniaxial magnetic anisotropy (IMA) ring within a perpendicular magnetic anisotropy (PMA) background, to create and stabilize skyrmions and merons. This geometric and anisotropic mechanism functions without chiral interactions, broadening the range of applicable materials to include centrosymmetric ones lacking DMI. By leveraging magnetic anisotropy and spatial patterning, tailored skyrmion lattices and non-collinear magnetic orders can be achieved.

In this letter, we strategically embed IMA rings within PMA boundaries, thereby transforming the role of geometric confinement from passive stabilization to active symmetry breaking. This approach yields a rich phase space of skyrmions, anti-skyrmions, and their fractional-meron counterparts, each exhibiting tunable in-plane symmetry via anisotropy modulation. Our work establishes spatial anisotropy engineering as a powerful, materials-generalizable method to topological spintronics, liberating skyrmionics from the constraints of chiral interactions.

These nanodisks, based on ferromagnet/heavy-metal heterostructures, offer a controllable platform for experimental verification. The DMI‑free configurations are specifically linked to FeCo or CoFeB/MgO systems with negligible DMI, while CoFeB/Pt remains an option for DMI‑present studies in the nanodisks. These structures, readily achievable via electron-beam-induced deposition (EBID) in a high-resolution dual-beam scanning electron microscope (SEM), as demonstrated in prior work \cite{dobro2020}, enable fine tuning of their magnetic properties through interface engineering \cite{vij2020, 1liu2021}. This interface engineering enables the manipulation of parameters such as the DMI and interlayer exchange coupling, thereby controlling the nanodisk's magnetic behavior. Furthermore, the strong perpendicular magnetic anisotropy (PMA), a crucial requirement for many spintronic applications and a key factor in our theoretical model, can be effectively induced in Fe/III-V nitride systems and ultrathin Co films \cite{yuGiant2018, liuStrong2017}. The well-established methods for achieving PMA in these materials provide a solid foundation for creating the desired magnetic environment within the nanodisks.

% \section{Theoretical model and Methodology}
\begin{figure}[h!]
\centering{}\includegraphics[width=0.8\linewidth]{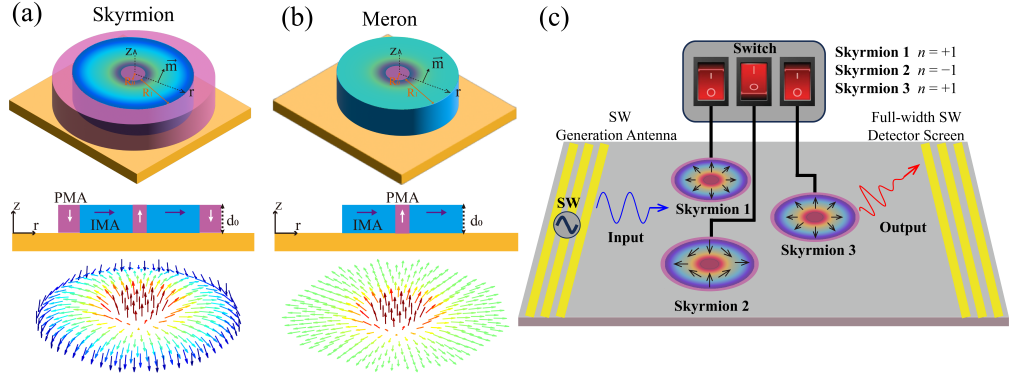}
\caption{Schematic model of the proposed non-chiral magnetic nanodisks by design, with radius $R=250$ nm and thickness $d_0$: (a) a skyrmion within an IMA nanoring of width $R_1=200$ nm, with strong PMA materials in its center of radius $R_0=50$ nm, and beyond the outer edge, and (b) a meron within an IMA nanoring of the same size with PMA material only in its center. (c) Schematic of SW control using anisotropy-engineered skyrmions. 3D spin structures are shown in (a) and (b), respectively.} 
\label{model}
\end{figure}

We consider two-dimensional (2D) magnetic nanodisks incorporating spatially engineered magnetic anisotropy, as illustrated in Figure \ref{model}. Two distinct configurations are examined: (i) a nanodisk consisting of an uniaxial IMA ring (radius $R_1$) surrounded by strong PMA, both at the center (radius $R_0$) and beyond the outer edge in Figure \ref{model}(a); and (ii) a nanodisk with PMA confined solely to the central region, while the surrounding ring exhibits IMA in Figure \ref{model}(b). Figure~\ref{model}(c) shows a proof-of-concept spin-wave control setup using anisotropy-engineered skyrmions. A microwave antenna on the left excites a spin wave (SW) traveling in the $x$ direction that interacts with skyrmions of vorticity $n=+1$ or $n=-1$. The reversal of the in-plane magnetization, depending on the sign of $n$, causes asymmetric spin-wave scattering in the transverse direction. Thus, the vorticity $n$ can be used to steer, focus, or diverge the transmitted spin wave.

Within the Landau-Lifshitz continuum approximation, the magnetization dynamics are described by the normalized vector field $\boldsymbol{m} = \boldsymbol{M}/M_s$, where $M_s$ is the saturation magnetization. In polar coordinates $(r,\phi)$, the magnetization is parameterized by the polar angle $\theta(r,\phi)$ measured from the $z$-axis and the azimuthal angle $\Phi(r,\phi)$:
\begin{equation}
\boldsymbol{m} = (\sin\theta\cos\Phi,\; \sin\theta\sin\Phi,\; \cos\theta).
\end{equation}
The azimuthal function takes the form $\Phi = n\phi + \gamma$, where $n$ is the vorticity (winding number) and $\gamma$ the helicity (initial phase) \cite{nagaosa_2013}. 

The equilibrium magnetization configuration is obtained by minimizing the total magnetic energy functional:
\begin{equation}
E[\theta] = d_0\int_0^{R}\int_0^{2\pi} \Big(e_{\text{ex}} + e_{\text{DM}} + e_{\text{ani}}+e_{\text{dem}}\Big)\, r\, dr d\phi, \label{Etot}
\end{equation}
where $d_0$ is the nanodisk thickness and $R = R_0 + R_1$ its total radius. The energy densities correspond to Heisenberg exchange $e_{\text{ex}}$, interfacial DMI $e_{\text{DM}}$ if we consider, and magnetic anisotropy $e_{\text{ani}}$, with associated material constants $A$, $D$, and $K$, respectively \cite{dongTuning2023, fu2025}. Moreover, the simplified demagnetization energy density $e_{\text{dem}}=1/2\mu_0(M_sm_z)^2$ for a thin disk is considered with the magnetic permeability $\mu_0$ in a vacuum. In the polar coordinates, as shown in Figure \ref{model}, the exchange energy density is
$e_{\text{ex}} = A\left[ \left(\frac{\partial\theta}{\partial r}\right)^2 + \left(\frac{1}{r}\frac{\partial\theta}{\partial\phi}\right)^2 + \left(\frac{n\sin\theta}{r}\right)^2 \right]$.
The interfacial DMI energy density, arising from inversion symmetry breaking at heterointerfaces \cite{yang267210}, takes the form 
$e_{\text{DM}} = -D(\cos\phi\cos\Phi + \sin\phi\sin\Phi)\left(\frac{\partial\theta}{\partial r} + \frac{n}{r}\cos\theta\sin\theta\right)
+ D\Big(\sin\phi\cos\Phi - \cos\phi\sin\Phi \Big)\frac{1}{r}\frac{\partial\theta}{\partial\phi}$. The anisotropy energy density for IMA along the $x$-direction ($\boldsymbol{\hat{e}} = (1,0,0)$) is $e_{\text{ani}} = -K(\boldsymbol{m}\cdot\boldsymbol{\hat{e}})^2 = -K(\sin\theta\cos\Phi)^2
$.

The analytical solutions correspond to constrained minima within the subspace of fixed $(n, \gamma)$ via a constrained variational approach, and their primary role is to identify candidate spin textures by minimizing the energy subject to the Euler-Lagrange equation and to reveal the physical trends induced by the competing anisotropies. The Euler-Lagrange equation for the energy functional \eqref{Etot} is derived from the condition $\frac{\partial L}{\partial\theta} - \sum_{i=r,\phi}\frac{d}{dx_i}\frac{\partial L}{\partial\theta'_{x_i}} = 0$, where the Lagrangian density $L = (e_{\text{ex}} + e_{\text{DM}} + e_{\text{ani}}+e_{\text{dem}})r$ \cite{Kiel2018, fu2025}. After algebraic manipulation, the governing partial differential equation for $\theta(r,\phi)$ becomes

{\small
\begin{equation}
\begin{aligned}
r^2\frac{\partial^2\theta}{\partial r^2} &+ \frac{\partial^2\theta}{\partial\phi^2} + r\frac{\partial\theta}{\partial r} - \left(\frac{n^2}{2} - \frac{r^2}{2R_K^2}\cos^2\Phi-\frac{r^2}{2R_s^2} \right)\sin2\theta \\
&- \frac{n r}{R_D}\sin^2\theta\Big(\cos\phi\cos\Phi + \sin\phi\sin\Phi\Big) = 0, \label{Eulag}
\end{aligned}
\end{equation}}
where the characteristic length scales $R_K = \sqrt{A/K}$, $R_D = A/D$ and $R_S = \sqrt{2A/\mu_0M_s^2}$ govern the anisotropy, DMI and demagnetization-dominated regimes, respectively.

Equation \eqref{Eulag} is solved numerically, subject to Dirichlet boundary conditions imposed by the surrounding PMA regions \cite{raftrey2021, fu2025}. For the configuration in Figure \ref{model}(a), the boundary conditions are $\theta(R_0,\phi) = 0$ (PMA at center) and $\theta(R,\phi) = \pi$ (PMA at outer edge). For Figure \ref{model}(b), we impose $\theta(R_0,\phi) = 0$ (central PMA) and $\theta(R,\phi) = \pi/2$ (outer IMA). These boundary conditions reflect the transfer of perpendicular anisotropy via exchange coupling across interfaces \cite{dieny2017}. 

The topological charge $Q$ characterizing the resulting spin textures is computed as
\begin{equation}
\begin{aligned}
Q &=\frac{1}{4\pi}\iint\mathrm{d}^2\mathbf{r}\left[\mathbf{m}\cdot\left(\frac{\partial\mathbf{m}}{\partial x}\times\frac{\partial\mathbf{m}}{\partial y}\right)\right] \\&= \frac{1}{4\pi}\int_0^R\int_0^{2\pi} \Big(n\sin\theta\,\frac{\partial\theta}{\partial r} \Big)\, dr d\phi, 
\end{aligned}\label{charge}
\end{equation}
which distinguishes skyrmions ($Q = \pm 1$) from merons ($Q = \pm 0.5$). The net normalized magnetization components are obtained via spatial averaging as 
\begin{equation}
(M_x, M_y, M_z) = \frac{1}{\mathcal{S}}\int_0^R\int_0^{2\pi} (m_x, m_y, m_z)\, r\, dr d\phi, \label{moment}
\end{equation}
where $\mathcal{S} = \pi R^2$ is the nanodisk area \cite{Gobel2020mqd}. These quantities enable quantitative analysis of symmetry breaking and its impact on macroscopic magnetic properties.

%\section{Resulsts and discussions}
We present the magnetic nanodisks with the realized spin topological structures in Figure \ref{model}(a) and (b), respectively. These configurations stem from the synergistic interplay between the IMA in the chiral nanoring and the boundary confinement imposed by PMA at the nanodisk perimeters. While the chiral DMI, when present, promotes the formation of non-trivial topologies, the IMA intrinsically breaks the in-plane axial symmetry, and the PMA-induced boundary confinement ensures the stability of spin textures. The precise engineering of nanoscale magnetic anisotropy enables the design of complex non-collinear spin textures beyond the conventional axisymmetric skyrmion paradigm \cite{dongTuning2023, wuSize2021}, offering deterministic control over both topology and internal symmetry.

\begin{figure}[h!]
\centering{}\includegraphics[width=1.0\columnwidth]{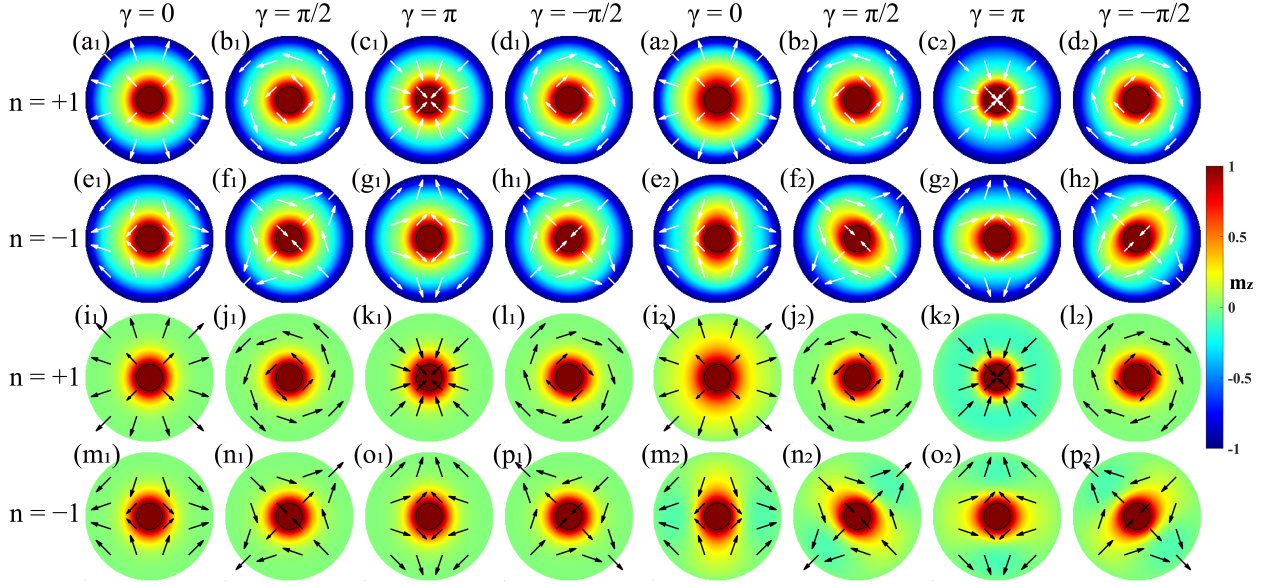}
\caption{The in-plane spin configurations of skyrmions stabilized by constrained analytical solutions: (a$_1$-h$_1$) for skyrmions and anti-skyrmions, (i$_1$-p$_1$) for merons and anti-merons without DMI; (a$_2$-h$_2$) for skyrmions and anti-skyrmions, (i$_2$-p$_2$) for merons and anti-merons with DMI. The used parameters are $A=16$ pJ/m, $K_{\text{PMA}}=3.6$ MJ/m$^3$ for PMA value, and the indicated IMA values of $K_{\text{IMA}}=5$ KJ/m$^3$ and $D=0.4$ mJ/m$^2$. }
\label{spin-noD}
\end{figure}

We now focus on the central finding of this work: skyrmions and merons can be stabilized even in the complete absence of chiral interaction ($D=0$). As shown in Figure \ref{spin-noD}, the nanodisk designs in Figure \ref{model}, even without any chiral interaction, a striking departure from conventional skyrmionics. The emergence of topologically nontrivial states arises solely from spatially competing anisotropies: the IMA within the nanoring favors in-plane alignment, whereas the strong PMA at the boundaries enforces a perpendicular orientation \cite{fu2025, dong2024}. Under appropriate boundary conditions and nanoscale dimensions, this competition drives the formation of helical skyrmion-like states, demonstrating that geometric confinement alone can replace chiral exchange interactions as a mechanism for stabilization. These findings establish that spatially engineered nanostructures provide a robust pathway to realizing skyrmionic states in conventional, non-chiral magnetic materials, thereby offering a practical route for chiral spin textures without requiring DMI interfaces.

\begin{figure}[h!]
\centering
\includegraphics[width=0.8\columnwidth]{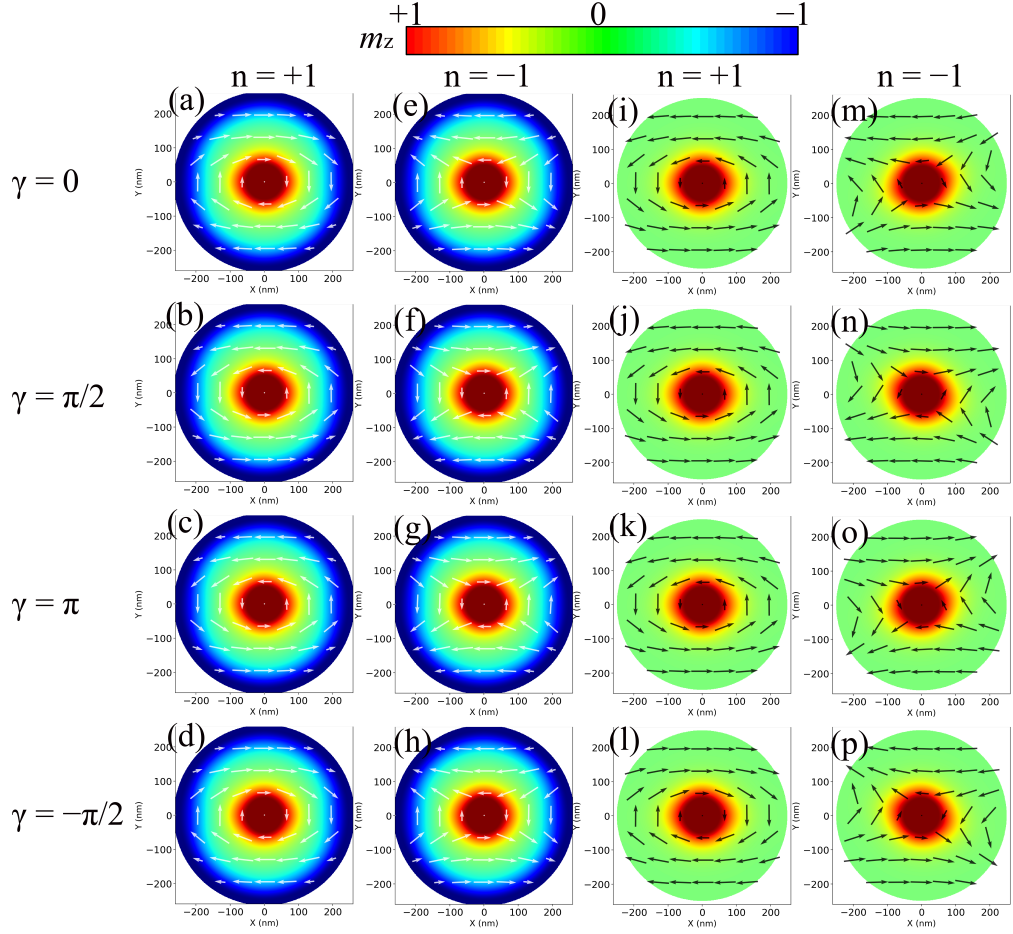}
\caption{The micromagnetic simulated in-plane spin configurations of skyrmions (a-h) and merons (i-p) in the absence of DMI ($D=0$) with dipolar interactions, respectively. The simulation parameters are $A=16$ pJ/m, $K_{\text{PMA}}=3.6$ MJ/m$^3$ for PMA value, and the indicated IMA values of $K_{\text{IMA}}=5$ KJ/m$^3$. }
\label{spin-D0-sim}
\end{figure}
\begin{figure}[h!]
\centering
\includegraphics[width=0.8\columnwidth]{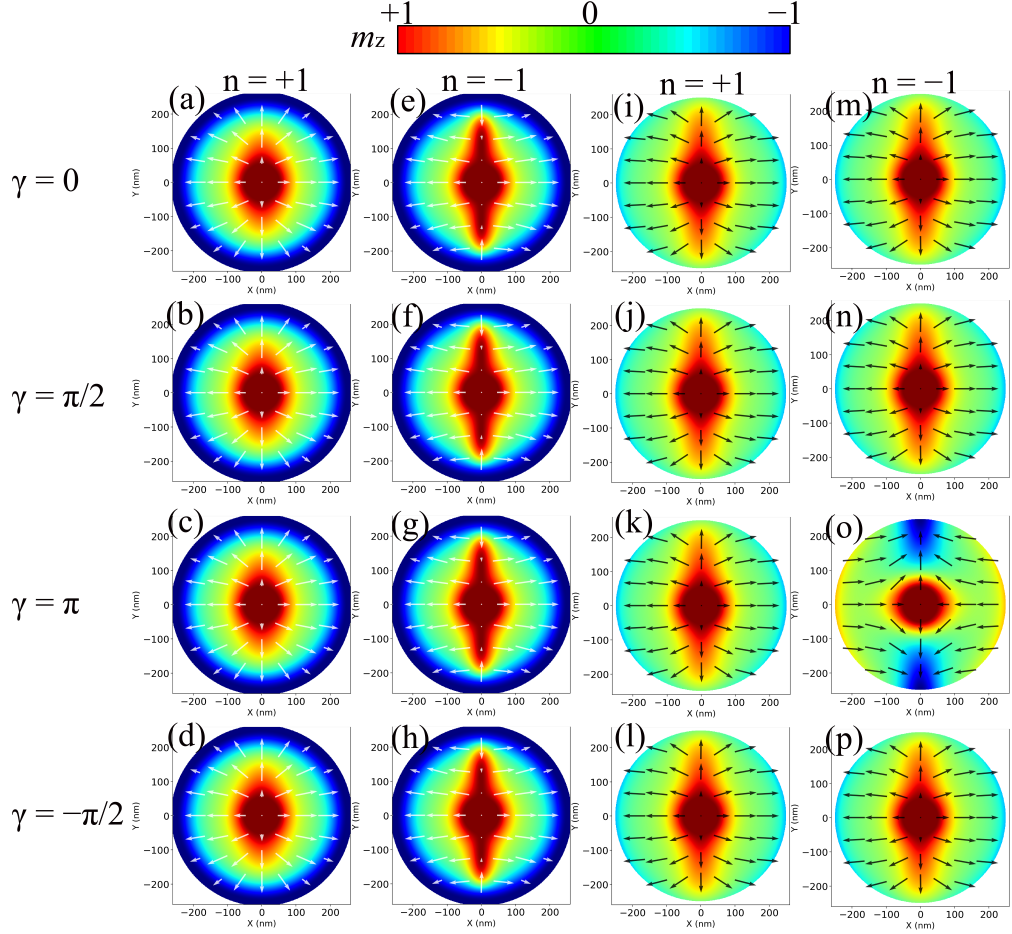}
\caption{The micromagnetic simulated in-plane spin configurations of skyrmions (a-h) and merons (i-p) in the presence of DMI with dipolar interactions, respectively. The simulation parameters are $A=16$ pJ/m, $K_{\text{PMA}}=3.6$ MJ/m$^3$ for PMA value, and the indicated IMA values of $K_{\text{IMA}}=5$ KJ/m$^3$ and $D=0.4$ mJ/m$^2$.}
\label{spin-D-sim}
\end{figure}

We have performed unconstrained micromagnetic simulations using Mumax3 to provide quantitative validation, with sufficiently long relaxation times for all the spin textures shown in Figure \ref{spin-D0-sim} for $D=0$ and Figure \ref{spin-D-sim} for $D=0.4$ mJ/m$^2$ (see Supporting Information S1 for details) \cite{Leliaert_2018}, including magnetic dipolar interactions. $n$ and $\gamma$ are only used as initial configurations for simulations. The simulated configurations, with and without DMI, and the theoretical spin configuration for comparison are presented in Figures S1-S4, respectively. This indicates that, in the absence of chiral DMI, skyrmions and merons can be realized simply by designing spatially anisotropic structures. However, the simulation results indicate that in the presence of DMI, the DMI induces the skyrmions and merons to relax into the same Néel type, while the antiskyrmions relax into the same anti-type; the vast majority of antimerons relax into Néel-type merons, with only a very small number retaining their antiskyrmion configuration, as shown in Figure \ref{spin-D-sim}. In the absence of DMI, the dipole interaction causes skyrmions and merons to relax into clockwise and counterclockwise Bloch configurations, while antimerons and antiskyrmions also relax into two types of anti-configurations. Moreover, spatially engineered anisotropy can produce meron-like textures, but DMI breaks the helicity degeneracy, thereby selecting preferred chiral configurations.

These simulation results in Figures \ref{spin-D0-sim} and \ref{spin-D-sim} show that the anisotropy-engineered nanodisks can support skyrmion-like and meron-like textures in the absence of DMI. However, the dipolar interaction can lift the helicity degeneracy and favor Bloch-like helicities, consistent with the experimental observations in centrosymmetric magnets \cite{yuSkyrmions2012}. Spatially engineered anisotropy can stabilize topological textures without DMI and can tailor their shape, size, and net magnetization. The helicity degeneracy is exact only in constrained theoretical model without DMI and is lifted when magnetostatics is included. Thus, the helicity degeneracy of the constrained theoretical model should be regarded as an idealized limit rather than a fully free degree of freedom. Moreover, the patterned anisotropy modifies the texture profile and produces a change in the spatially averaged magnetization. Detailed $M_z$ profiles are shown in Figure S5 of Supporting Information.

\begin{figure}[h!]
\centering{}\includegraphics[width=0.80\columnwidth]{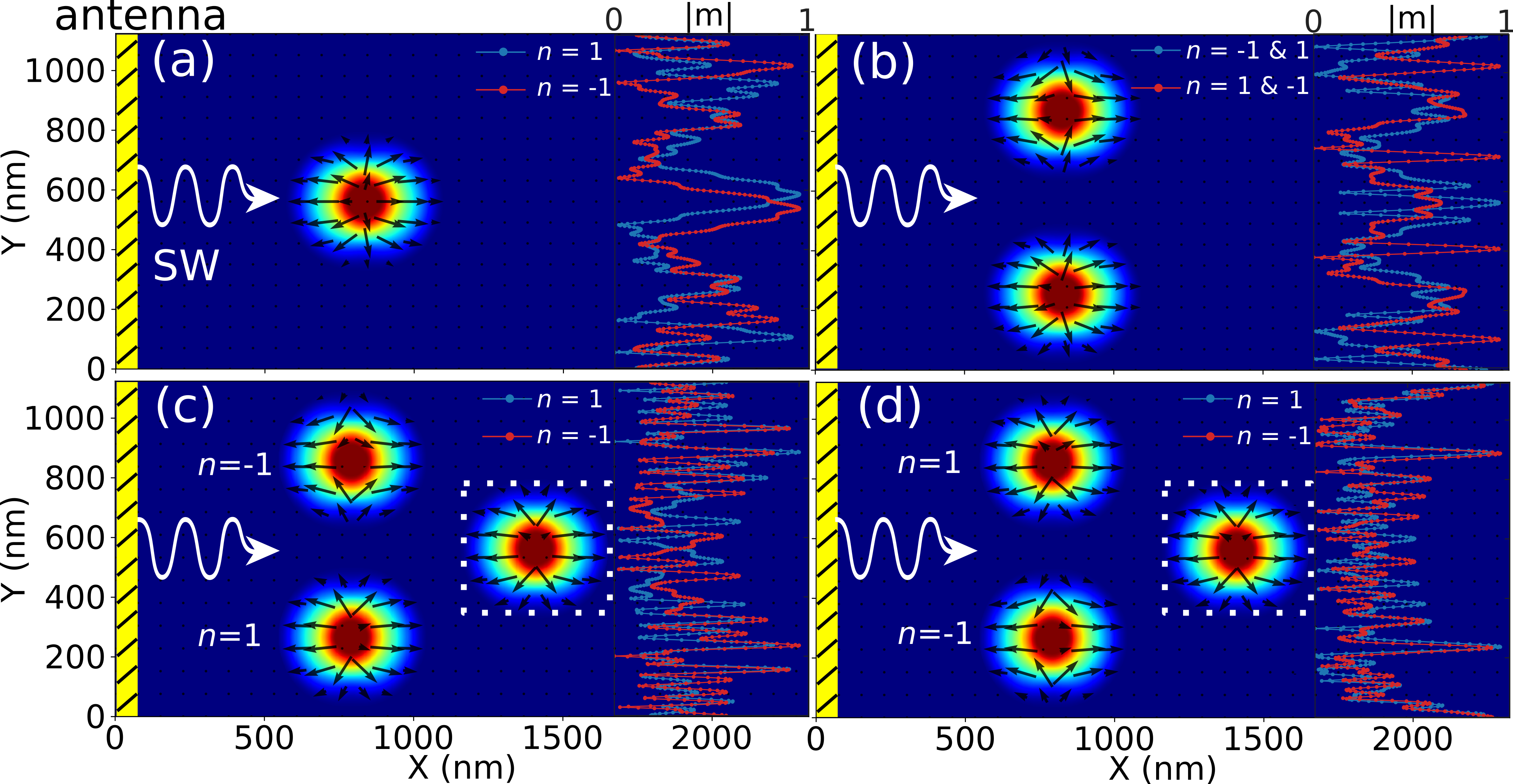}
\caption{Vorticity-controlled spin-wave propagation through anisotropy-engineered skyrmion arrays. A microwave antenna at the left boundary excites a 300 GHz spin wave propagating along the $x$ direction. (a) A single skyrmion with $n=+1$ or $n=-1$ produces opposite transverse deflections of the transmitted wave. (b) A pair of skyrmions with opposite vorticities redistributes the transmitted amplitude toward or away from the central channel according to their vertical sequence. (c) A three-skyrmion arrangement focuses the signal toward the central channel and then deflects it according to the vorticity of the downstream skyrmion. (d) The complementary arrangement produces an off-center, divergent transmission profile. }
\label{sw}
\end{figure}

To demonstrate a concrete functionality of the anisotropy-engineered textures, we performed time-dependent Mumax3 simulations of spin-wave propagation through skyrmion arrays in the DMI-free exchange-anisotropy model. As illustrated in Figure \ref{sw}, we investigate the engineering of spin-wave wavefronts utilizing the chirality-dependent scattering properties of magnetic skyrmions. While the fundamental scattering of a SW by a single, isolated skyrmion is well-established in the literature \cite{markus2014}, demonstrated in Figure \ref{sw}(a), where distinct forward-scattering intensity profiles emerge for chiralities $n=+1$ and $n=-1$, our work explicitly exploits this asymmetry to design tunable magnonic elements. By positioning a transverse pair of skyrmions with specific chiral combinations, as shown in Figure \ref{sw}(b), the composite system can be tailored to function as a SW lens, effectively focusing or diverging the incident wave based on the chosen topological states. Furthermore, we demonstrate advanced dynamic control by introducing a third skyrmion downstream to create a more complex interference pattern. As shown in Figure \ref{sw}(c), this specific three-skyrmion configuration acts as a directional router; depending on the chirality of this third skyrmion, the majority of the SW can be selectively scattered either upward or downward. Conversely, modifying the chiral configuration of the initial upstream pair, as depicted in Figure \ref{sw}(d), yields an entirely different type of complex scattering behavior and output profile. Consequently, this multi-skyrmion architecture provides a highly tunable and scalable mechanism for directional SW routing and reconfigurable magnonic logic operations.

Furthermore, our model assumes a perfectly homogeneous magnetic medium, real samples inevitably contain quenched disorder such as grain boundaries, interface roughness, and local variations of magnetic anisotropy \cite{R035005, P214403}. Based on theoretical analyses, such disorder can pin skyrmions and merons, effectively enhancing their stability against perturbations. Compared to natural skyrmion lattices, the skyrmion lattice structure we designed can be tailored, and the skyrmion states do not move during spin-wave scattering. The present relaxation calculations establish very low temperature configurations but do not provide a quantitative thermal lifetime. Such a lifetime requires a transition-path calculation, for example by a geodesic nudged elastic band or related minimum-energy-path method, and/or finite-temperature stochastic Landau-Lifshitz-Gilbert simulations \cite{Pavel2015, Hahn_2019}. These analyses are beyond the scope of the present work.

In conclusion, we have shown that spatially engineered magnetic anisotropy can stabilize skyrmion-like and meron-like spin textures in magnetic nanodisks without requiring chiral DMI. The competing perpendicular and in-plane anisotropies impose non-collinear magnetic configurations through the patterned anisotropy landscape. In the idealized exchange--anisotropy model, different trial vorticities and helicities can be constructed, whereas unconstrained micromagnetic simulations show that DMI and magnetostatic interactions select preferred relaxed chiral configurations. As a proof-of-principle magnonic functionality, we further demonstrate that the vorticity and arrangement of anisotropy-engineered skyrmions control the transverse redistribution of a transmitted spin wave. The resulting steering, central-channel focusing, and off-center divergence arise from the combination of vorticity-dependent skew-scattering events. These results establish patterned anisotropy as a route to DMI-free topological textures and to designable spin-wave amplitude profiles in nanoscale magnetic structures.

\section{Notes}
The authors declare no competing financial interests.

\section{Acknowledgments}  
This work is funded by the Science and Technology Program of Xuzhou (KC25001) and supported by the National Natural Science Foundation of China (Grant No. 12374079), the Research Foundation-Flanders (FWO-Vlaanderen), and the Special Research Funds of the University of Antwerp (BOF-UA). Z. H. Tao gratefully acknowledges support from the China Scholarship Council.

\textit{}\section{Supporting Information Available}
In the Supporting Information, we provide details of the simulation parameters used. 
The skyrmions and merons 3D structures, along with the simulation method and details, are presented using Mumax3 simulations and compared with our theoretical results. The magnetization profile $m_z$ and the control of the spin wave propagation are shown.

\bibliography{achemso-demo}

\end{document}